\documentclass[aps,prab,reprint,amsmath,amssymb,nofootinbib]{revtex4-2}

\usepackage{bm}
\usepackage{graphicx}
\usepackage{tikz}
\usetikzlibrary{arrows.meta,calc,decorations.pathreplacing}
\usepackage{hyperref}
\hypersetup{
 hidelinks,
 pdfauthor={I. Zagorodnov},
 pdftitle={Indirect Integration of Longitudinal and Transverse Wake Potentials}
}

\newcommand{\rp}{\bm r_{\!\perp}}
\newcommand{\gradp}{\nabla_{\!\perp}}
\newcommand{\Deltap}{\Delta_{\!\perp}}
\newcommand{\divp}{\nabla_{\!\perp}\!\cdot}
\newcommand{\curlp}{(\nabla_{\!\perp}\!\times\,)_{z}}
\newcommand{\ez}{\bm e_z}
\newcommand{\E}{\bm E}
\newcommand{\B}{\bm B}
\newcommand{\Fp}{\bm F_{\!\perp}}
\newcommand{\Kp}{\bm K_{\!\perp}}
\newcommand{\Wp}{\bm W_{\!\perp}}
\newcommand{\dd}{\,\mathrm d}
\newcommand{\ii}{\mathrm i}

\begin{document}

\title{Indirect Integration of Longitudinal and Transverse Wake Potentials for Unequal Beam Pipes and Arbitrary Beam Velocity}

\author{Igor Zagorodnov}
\email[Corresponding author: ]{igor.zagorodnov@desy.de}
\author{Dmitry Bazyl}
\affiliation{Deutsches Elektronen-Synchrotron DESY, Notkestrasse 85, 22607 Hamburg, Germany}

\date{\today}

\begin{abstract}
Indirect integration replaces the long uniform beam-pipe parts of a wakefield calculation by field
problems in the pipe cross sections.  The earlier ultrarelativistic methods were developed mainly for
the longitudinal wake, while the transverse wake was usually obtained from the Panofsky--Wenzel
theorem.  In this paper we derive indirect formulas which complete a transverse Lorentz-force
integral already accumulated in a time-domain calculation.  At $\beta=1$, each semi-infinite tail is
found from a Dirichlet Poisson problem driven by $E_z$ and a Neumann Poisson problem driven by
$cB_z$.  These two problems describe the TM and TE contributions, respectively.  We obtain both a
fixed-time moving-window representation and a fixed-plane time-history representation for equal or
unequal input and output pipes.  We then extend the method to a rigid bunch moving with constant
velocity $0<\beta c<c$.  In this case the longitudinal correction satisfies an anisotropic elliptic
equation in $(x,y,s)$ and gives an additional source for the transverse TM problem.  For a unified two-port finite-reference convention, the complete fields, including space charge,
are integrated directly between two fixed integration planes for both equal and unequal pipes.  The two
semi-infinite tails are calculated after subtraction of the stationary field of the corresponding
pipe.  When the two stationary pipe fields are identical, this finite-reference quantity differs
from the globally subtracted scattered-field wake only by the known pure-pipe contribution over the
port separation.  The resulting Panofsky--Wenzel relation contains the difference of the stationary
transverse electric fields in the two pipes.  Finally, the transverse method is tested for an
unequal rectangular step-out at both $\beta=1$ and $\beta=0.8$.  The finite-velocity result
includes the anisotropic longitudinal solve and the additional transverse TM source, and it agrees
with the longitudinal--transverse relation including the unequal-pipe boundary term.
\end{abstract}

\maketitle

\section{Introduction}

The direct calculation of wake potentials becomes inefficient when the fields scattered by a
localized structure need a long distance to catch up with a relativistic witness.  Indirect methods
replace the integration in the long uniform pipes by field problems in one or more cross sections.
For a time-domain solver it is important to distinguish between the force already integrated
directly and the missing semi-infinite tail.  A reconstruction of only the longitudinal wake,
followed by a global Panofsky--Wenzel integration, does not complete a transverse Lorentz-force
integral accumulated independently.

Several earlier results are particularly relevant to the present work. For cavity-like axisymmetric structures, Weiland showed that the improper longitudinal wake integral can be reduced to a finite integral over the cavity gap~\cite{Weiland1983}. Napoly, Chin, and Zotter subsequently expressed the longitudinal and transverse wake potentials of arbitrary multipole order in terms of field integrals evaluated along a deformable longitudinal contour~\cite{NapolyChinZotter1993}. A one-sided treatment of the downstream tail was later introduced in the moving-mesh ECHO algorithm by Zagorodnov, Schuhmann, and Weiland~\cite{ZagorodnovSchuhmannWeiland2003}. In that formulation, only the scattered longitudinal field component, $E_z^{\mathrm{sc}}$, is integrated along the straight witness trajectory, whereas the remaining field components enter through a correction evaluated at the downstream end of the computational domain. This construction also permits the treatment of structures with unequal input and output pipes. The transverse wake is then obtained from the longitudinal wake by means of the Panofsky--Wenzel theorem~\cite{PanofskyWenzel1956}.

The general three-dimensional longitudinal problem was treated in two different ways in 2006.
Zagorodnov used an expansion in output-waveguide modes and obtained both a fixed-time moving-window
Poisson formulation and a fixed-plane time-history formulation~\cite{Zagorodnov2006}.  Henke and
Bruns derived a contour deformation for a fixed computational domain and a general three-dimensional
structure~\cite{HenkeBruns2006}.   Shobuda, Chin, and Takata later derived an explicit
generalized Napoly formula for axisymmetric structures with unequal entrance and exit radii and
implemented it in ABCI~\cite{ShobudaChinTakata2008}.  

These methods give the longitudinal wake and,
through the Panofsky--Wenzel relation, the transverse wake.  However, they do not directly complete
an unfinished transverse Lorentz-force integral in an arbitrary three-dimensional output pipe from
the longitudinal field components.

At finite velocity the stationary space-charge field in a uniform pipe cannot be omitted.  It has a
nonzero longitudinal component and a nonzero transverse Lorentz force.  Hence, its complete-field
integral over a semi-infinite pipe is not a localized geometric wake.  If the two asymptotic pipes
have the same stationary field, this nonlocalized contribution can be subtracted globally and a
port-independent scattered-field wake can be defined.  For unequal pipes there is no unique global
pipe reference through the device.  We therefore adopt one two-port finite-reference convention for
both equal and unequal pipes: the complete device field is retained between the fixed integration
planes, while the stationary field of the matching pipe is subtracted only in each uniform tail.
The equal-pipe scattered-field wake is recovered from this quantity by an explicit pure-pipe
correction.  The endpoint conditions for the Panofsky--Wenzel theorem are
known~\cite{VaganianHenke1995} and will be used below.

In this paper we reconstruct directly the missing transverse impulse.  At $\beta=1$, its transverse
divergence and curl are determined by $E_z$ and $B_z$ in a pipe cross section.  A Hodge decomposition
then gives a Dirichlet Poisson problem for the TM part and a Neumann Poisson problem for the TE part.
We derive the fixed-time moving-endpoint and the fixed-plane time-history forms.  We then extend the
same method to a prescribed rigid bunch with $0<\beta<1$.  In this case the TM problem contains an
additional $\gamma^{-2}\partial_s$ term from the longitudinal tail correction.  Finally, we give
a unified two-port formulation for equal or unequal input and output pipes without introducing a
reference-switching plane inside the structure.

The numerical examples in Sec.~\ref{sec:examples} use the same unequal rectangular step-out at
$\beta=1$ and at the strictly finite velocity $\beta=0.8$.  The latter exercises the coupled
stationary pipe field, the modified Helmholtz terminal problems, and the
$\gamma^{-2}\partial_s$ contribution to the transverse TM reconstruction.

The paper is organized as follows.  In Sec.~\ref{sec:definitions} we introduce the geometry,
notation, and wake convention.  In Sec.~\ref{sec:known-beta1} we review the longitudinal methods at
$\beta=1$.  In Sec.~\ref{sec:transverse-beta1} we derive the direct transverse reconstruction.
Section~\ref{sec:finite-beta} extends the result to finite velocity and develops the unified
two-port convention for equal or unequal pipes and describes its numerical solution.  In
Sec.~\ref{sec:PW} we derive the corresponding Panofsky--Wenzel relations.  In
Sec.~\ref{sec:examples} we test the transverse method for an unequal rectangular step-out.
Appendix~\ref{app:one-sided} collects the one-sided finite-velocity representations, their
conversion identities, and their Panofsky--Wenzel consistency check.

\section{Geometry, coordinates, and wake convention}
\label{sec:definitions}

\begin{figure*}[t]
\centering
 \includegraphics[width=0.98\textwidth]{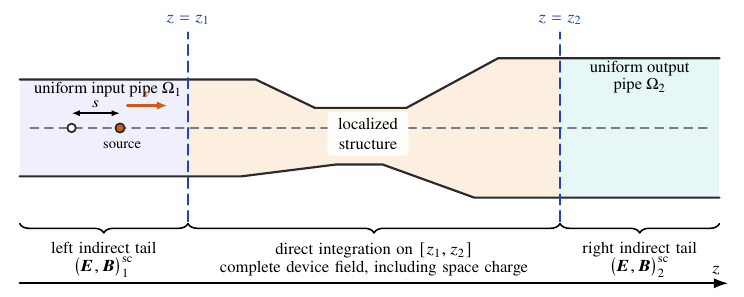}
\caption{Geometry and field convention for unequal beam pipes.  The complete device fields are
integrated directly only between the fixed port planes $z_1$ and $z_2$.  Each semi-infinite
indirect tail uses a local scattered field obtained by subtracting the stationary field of the
matching pipe.  No reference-switching interface is introduced inside the structure.}
\label{fig:unequal-pipe-schematic}
\end{figure*}

In this paper we consider a localized vacuum structure with perfectly conducting walls.  It is connected to a
uniform input pipe $\Omega_1\times(-\infty,z_1]$ and a uniform output pipe
$\Omega_2\times[z_2,\infty)$.  The transverse sections $\Omega_1$ and $\Omega_2$ may have
different sizes and shapes.  Source and witness trajectories are straight, parallel to the $z$ axis,
and contained in the common aperture.  Both particles move with the prescribed constant velocity
$v=\beta c$.  Figure~\ref{fig:unequal-pipe-schematic} shows the geometry, the two fixed port
planes, and the field convention used below.

The source reference point is at $z=vt$.  A witness at distance $s$ behind it samples a field at
\begin{equation}
 t=t(z,s)=\frac{z+s}{v},
 \qquad
 \overline{f}(\rp,z,s)=f\!\left(\rp,z,\frac{z+s}{v}\right).
 \label{eq:pullback}
\end{equation}
Thus $s>0$ denotes a trailing witness.  Along the sampled trajectory,
\begin{equation}
 \frac{\dd \overline f}{\dd z}
 =\left(\partial_z+\frac{1}{v}\partial_t\right)f,
 \qquad
 \partial_s\overline f=\frac{1}{v}\partial_t f.
 \label{eq:chain}
\end{equation}
We use
\begin{equation}
 D_{\pm}=\partial_z\pm v^{-1}\partial_t,
 \qquad
 \gamma=(1-\beta^2)^{-1/2}.
 \label{eq:Dpm}
\end{equation}

The transverse Lorentz force per unit witness charge is
\begin{equation}
 \Fp=\E_\perp+v\ez\times\B_\perp.
 \label{eq:force}
\end{equation}
For a drive-bunch charge $Q$, the usual formal wake definitions are
\begin{align}
 QW_\parallel(\rp,s)&=-\int_{-\infty}^{\infty}\overline E_z(\rp,z,s)\dd z,
 \label{eq:formal-long-wake}\\
 Q\Wp(\rp,s)&=\int_{-\infty}^{\infty}\overline\Fp(\rp,z,s)\dd z.
 \label{eq:formal-trans-wake}
\end{align}
At $\beta=1$ these integrals are well defined with the usual outgoing-wave or limiting-absorption
prescription.  For $\beta<1$, the full-line integrals of the complete field diverge because they contain
the stationary space-charge field.  The corresponding finite-velocity definitions are introduced in
Sec.~\ref{sec:finite-beta}.  The main convention of this paper is the two-port finite-reference wake
of Sec.~\ref{sec:two-port}, which is defined for both equal and unequal pipes.

\section{Known longitudinal indirect methods at \texorpdfstring{$\beta=1$}{beta=1}}
\label{sec:known-beta1}

We first recall a useful property of the stationary field of a rigid bunch in a uniform PEC
waveguide.  Let the superscript $\mathrm p$ denote this pure-pipe field and put $\zeta=z-ct$.  In the
Lorenz gauge its potentials can be chosen in the form
\begin{equation}
 \bm A^{\mathrm p}=c^{-1}V^{\mathrm p}\ez,
 \qquad
 \Deltap V^{\mathrm p}=-\frac{\rho(\rp,\zeta)}{\varepsilon_0},
 \qquad
 V^{\mathrm p}|_{\partial\Omega}=0.
 \label{eq:beta1-pipe-potentials}
\end{equation}
The shape of the cross section $\Omega$ is arbitrary.  Since
$\partial_tV^{\mathrm p}=-c\partial_\zeta V^{\mathrm p}$ and
$\partial_zV^{\mathrm p}=\partial_\zeta V^{\mathrm p}$, Eq.~\eqref{eq:beta1-pipe-potentials} gives
\begin{equation}
 \begin{aligned}
 E_z^{\mathrm p}
 &=-\partial_zV^{\mathrm p}-\partial_t A_z^{\mathrm p}=0,\\
 B_z^{\mathrm p}
 &=\ez\cdot\nabla\times(A_z^{\mathrm p}\ez)=0,\\
 H_z^{\mathrm p}&=\mu_0^{-1}B_z^{\mathrm p}=0.
 \end{aligned}
 \label{eq:beta1-pipe-longitudinal-zero}
\end{equation}
The transverse components satisfy
\begin{equation}
 \begin{aligned}
 \E_\perp^{\mathrm p}&=-\gradp V^{\mathrm p},\\
 \B_\perp^{\mathrm p}&=c^{-1}\ez\times\E_\perp^{\mathrm p},\\
 \Fp^{\mathrm p}&=\E_\perp^{\mathrm p}
  +c\ez\times\B_\perp^{\mathrm p}=0.
 \end{aligned}
 \label{eq:beta1-pipe-force-zero}
\end{equation}
Thus the stationary pipe field contributes neither $E_z$, $B_z$, nor the transverse Lorentz force
at $\beta=1$.  The complete field can therefore be used throughout Secs.~\ref{sec:known-beta1} and
\ref{sec:transverse-beta1}; no pipe-field subtraction is needed.

\subsection{Moving computational window}

Let a moving-window calculation be stopped at $t_0=z_0/c$, with $z_0-s$ lying in a uniform output
pipe.  The moving-window tail integral is
\begin{equation}
 u(\rp,s)=
 \int_{z_0-s}^{\infty}\overline E_z(\rp,z,s)\dd z.
 \label{eq:moving-tail}
\end{equation}
Equation~(21) of Ref.~\cite{Zagorodnov2006} is equivalently
\begin{align}
 \Deltap u
 &=D_-E_z(\rp,z_0-s,t_0),
 &u|_{\partial\Omega_2}&=0.
 \label{eq:Z-moving}
\end{align}
The complete longitudinal wake is the directly accumulated integral ending at the same moving point
plus $u$,
\begin{equation}
 QW_\parallel=-\int_{-\infty}^{z_0-s}\overline E_z\dd z
              -u.
 \label{eq:Z-moving-total}
\end{equation}
All right-hand sides are taken from one spatial snapshot at $t=t_0$, while different values of $s$
correspond to different slices $z=z_0-s$.  The result above shows why no reference field has to be
specified in the output pipe at $\beta=1$.

\subsection{Fixed computational domain}

We write the fixed-domain result in two-sided tail notation.  Put
$\sigma_1=-1$, $\sigma_2=+1$ and define
\begin{equation}
 w_1=\int_{-\infty}^{z_1}\overline E_{z,1}\dd z,
 \qquad
 w_2=\int_{z_2}^{\infty}\overline E_{z,2}\dd z.
 \label{eq:beta1-wtails}
\end{equation}
At $\beta=1$ both tails follow from independent cross-sectional problems,
\begin{align}
 \Deltap w_j
 &=\sigma_jD_-E_{z,j}(\rp,z_j,t_j),
 &w_j|_{\partial\Omega_j}&=0,
 \label{eq:beta1-two-Poisson}\\
 t_j&=(z_j+s)/c.\nonumber
\end{align}
For the rigid ultrarelativistic current, $J_z=c\rho(\rp,z-ct)$, the source term in the
longitudinal electric-field wave equation vanishes because
$\mu_0\partial_tJ_z+\varepsilon_0^{-1}\partial_z\rho=0$.  Thus the complete longitudinal field in
either uniform pipe obeys
\begin{equation}
 \Deltap E_z
 =-\left(\partial_z^2-c^{-2}\partial_t^2\right)E_z
 =-D_+D_-E_z.
 \label{eq:beta1-wave-factor}
\end{equation}
For any field component $g$, the pullback~\eqref{eq:pullback} satisfies
\[
 \overline{D_+g}(\rp,z,s)
 =\left(\partial_z+c^{-1}\partial_t\right)
   g\!\left(\rp,z,\frac{z+s}{c}\right)
 =\frac{\dd\overline g}{\dd z}.
\]
Since $D_+$ and $D_-$ commute, Eq.~\eqref{eq:beta1-wave-factor}, evaluated on the
synchronous trajectory, becomes
\[
 \Deltap\overline E_{z,j}
 =-\frac{\dd}{\dd z}
   \overline{D_-E_{z,j}},
 \qquad j=1,2.
\]
All improper integrals below are understood with the same outgoing-wave, or equivalently
limiting-absorption, prescription as the wake integral.  More explicitly, one may first introduce
an arbitrarily small positive absorption.  The outgoing fields then decay at the remote end of each
pipe, so the corresponding endpoint term vanishes and the transverse Laplacian can be interchanged
with the tail integral.  The zero-absorption limit is taken only after the following integrations.
For the output tail this gives
\begin{align*}
 \Deltap w_2
 &= -\int_{z_2}^{\infty}
      \frac{\dd}{\dd z}
      \overline{D_-E_{z,2}}\dd z \\
 &= -\left[
      \overline{D_-E_{z,2}}
      \right]_{z_2}^{\infty}
  =D_-E_{z,2}(\rp,z_2,t_2).
\end{align*}
For the input tail, the finite endpoint is the upper rather than the lower integration limit, and
therefore
\begin{align*}
 \Deltap w_1
 &= -\int_{-\infty}^{z_1}
      \frac{\dd}{\dd z}
      \overline{D_-E_{z,1}}\dd z \\
 &= -\left[
      \overline{D_-E_{z,1}}
      \right]_{-\infty}^{z_1}
  =-D_-E_{z,1}(\rp,z_1,t_1).
\end{align*}
These two endpoint evaluations explain the signs $\sigma_1=-1$ and $\sigma_2=+1$ in
Eq.~\eqref{eq:beta1-two-Poisson}.

On the longitudinal PEC wall, $\ez$ is tangential and therefore $E_z=0$.  It follows directly from
Eq.~\eqref{eq:beta1-wtails} that
$w_j|_{\partial\Omega_j}=0$, which supplies the Dirichlet boundary conditions in
Eq.~\eqref{eq:beta1-two-Poisson}.

It remains to relate the two tail solutions to the desired straight-line integral of the device
field.  Splitting that integral at the two fixed port planes gives the identity
\begin{align*}
 \int_{-\infty}^{\infty}\overline E_z\dd z
 &={}
 \int_{-\infty}^{z_1}\overline E_{z,1}\dd z
 +\int_{z_1}^{z_2}\overline E_z\dd z \\
 &\quad+\int_{z_2}^{\infty}\overline E_{z,2}\dd z.
\end{align*}
Using Eq.~\eqref{eq:beta1-wtails}, this identity becomes
\[
 \int_{-\infty}^{\infty}\overline E_z\dd z
 =w_1+\int_{z_1}^{z_2}\overline E_z\dd z+w_2.
\]
Together with the wake convention~\eqref{eq:formal-long-wake}, this yields
\begin{equation}
 QW_\parallel=-w_1-
 \int_{z_1}^{z_2}\overline E_z\dd z-w_2.
 \label{eq:beta1-fixed-total}
\end{equation}
Thus Eq.~\eqref{eq:beta1-fixed-total} is obtained by a direct partition of the full longitudinal
field integral, without deforming the integration path or introducing transverse endpoint
potentials.  No additional endpoint terms occur because $E_z^{\mathrm p}=0$.
Equations~\eqref{eq:Z-moving} and \eqref{eq:beta1-two-Poisson} are the moving-snapshot and fixed-plane
forms of the same elimination of the uniform waveguides.

\section{Transverse indirect integration at \texorpdfstring{$\beta=1$}{beta=1}}
\label{sec:transverse-beta1}

We now assume that the transverse force has already been integrated directly between the fixed
planes $z_1$ and $z_2$.  As shown in Sec.~\ref{sec:known-beta1}, the stationary pipe field has
$E_z^{\mathrm p}=B_z^{\mathrm p}=0$ and $\Fp^{\mathrm p}=0$.  We can therefore use the complete
field in the two tails.  From Maxwell's equations and Eq.~\eqref{eq:force}, with $v=c$, we obtain
the following identities.  In Cartesian components they read
\begin{equation}
 F_x=E_x-cB_y,
 \qquad
 F_y=E_y+cB_x.
 \label{eq:beta1-components}
\end{equation}
Gauss' law and the longitudinal component of the Amp\`ere--Maxwell law yield
\begin{align}
 \divp\Fp
 &=\partial_xE_x+\partial_yE_y
   -c(\partial_xB_y-\partial_yB_x)\nonumber\\
 &=\frac{\rho}{\varepsilon_0}-\partial_zE_z
   -c\mu_0J_z-c^{-1}\partial_tE_z\nonumber\\
 &=-\partial_zE_z-c^{-1}\partial_tE_z
   =-D_+E_z .
 \label{eq:beta1-div-derivation}
\end{align}
In the last step the charge and current terms cancel because $J_z=c\rho$ and
$\mu_0\varepsilon_0c^2=1$.
Likewise, Faraday's law and $\nabla\cdot\B=0$ give
\begin{align}
 \curlp\Fp
 &=\partial_xE_y-\partial_yE_x
   +c(\partial_xB_x+\partial_yB_y)\nonumber\\
 &=-\partial_tB_z-c\partial_zB_z
   =-cD_+B_z .
 \label{eq:beta1-curl-derivation}
\end{align}
On the witness trajectory $D_+=\dd/\dd z$.  Hence,
\begin{equation}
 \divp\overline\Fp
 =-\frac{\dd\overline E_z}{\dd z},
 \qquad
 \curlp\overline\Fp
 =-c\frac{\dd\overline B_z}{\dd z}.
 \label{eq:beta1-local}
\end{equation}
We define the two tail impulses as
\begin{align}
 \Kp^{(1)}&=\int_{-\infty}^{z_1}\overline{\bm F}_{\!\perp,1}\dd z,
 \label{eq:Ktail-left}\\
 \Kp^{(2)}&=\int_{z_2}^{\infty}\overline{\bm F}_{\!\perp,2}\dd z.
 \label{eq:Ktails}
\end{align}
Integration of Eq.~\eqref{eq:beta1-local} gives
\begin{align}
 \divp\Kp^{(j)}&=\sigma_jE_{z,j}(\rp,z_j,t_j),
 \label{eq:beta1-tail-div}\\
 \curlp\Kp^{(j)}&=\sigma_jcB_{z,j}(\rp,z_j,t_j).
 \label{eq:beta1-tail-divcurl}
\end{align}

On a longitudinal perfectly conducting wall, $E_z=E_\tau=0$ and $B_n=0$.  If
$\bm\tau=\ez\times\bm n$ is tangent to the cross-sectional boundary, then
$\Fp\cdot\bm\tau=E_\tau+cB_n=0$.  Hence, the tangential component of every tail impulse vanishes.
For a simply connected cross section we use the Hodge representation
\begin{equation}
 \Kp^{(j)}=\gradp\Phi_j+\ez\times\gradp\Psi_j
 \label{eq:Hodge}
\end{equation}
which is obtained from
\begin{align}
 \Deltap\Phi_j&=\sigma_jE_{z,j}(\rp,z_j,t_j),
 &\Phi_j|_{\partial\Omega_j}&=0,
 \label{eq:beta1-Phi}\\
 \Deltap\Psi_j&=\sigma_jcB_{z,j}(\rp,z_j,t_j),
 &\partial_n\Psi_j|_{\partial\Omega_j}&=0,
 \label{eq:beta1-Psi}\\
 &&\int_{\Omega_j}\Psi_j\dd A&=0.\nonumber
\end{align}
The divergence and curl of Eq.~\eqref{eq:Hodge} are $\Deltap\Phi_j$ and $\Deltap\Psi_j$,
respectively.  A constant Dirichlet value for $\Phi_j$ and the homogeneous Neumann condition for
$\Psi_j$ give $\Kp^{(j)}\cdot\bm\tau=0$.  We set the Dirichlet constant to zero and fix the
remaining additive constant in $\Psi_j$ by its zero mean.  The solution is then unique.
The Neumann compatibility condition is
$\int_{\Omega_j}B_{z,j}\dd A=0$; it follows from magnetic-flux conservation and the
absence of an externally imposed dc flux.  Equations~\eqref{eq:beta1-Phi} and
\eqref{eq:beta1-Psi} give the TM and TE tails, respectively.  The TE problem cannot in general be
omitted.  A TE mode has $E_z=0$, but it can contribute to the unfinished transverse-force integral.

Define the central directly integrated impulse by
\begin{equation}
 \Kp^{\mathrm{dir}}=\int_{z_1}^{z_2}\overline\Fp\dd z.
 \label{eq:beta1-direct}
\end{equation}
Finally, for equal or unequal pipes we obtain
\begin{equation}
 Q\Wp=\Kp^{(1)}+\Kp^{\mathrm{dir}}+\Kp^{(2)}.
 \label{eq:beta1-trans-final}
\end{equation}
When the pipe sections are unequal, the potentials for $j=1$ and $j=2$ are calculated on different
meshes.  Both impulses are then evaluated at the same physical witness position.

\section{Generalization to \texorpdfstring{$0<\beta<1$}{0 < beta < 1}}
\label{sec:finite-beta}

\subsection{Stationary field in a uniform pipe}

For finite $\gamma$ we use as a reference the stationary field of the same rigid bunch in the
corresponding pipe, and not its free-space field.  Superscript $\mathrm p$
denotes this stationary pipe field.  We write
$\zeta=z-vt$ and let
$\rho_j(\rp,\zeta)$ be the prescribed bunch density in pipe $j$.  The stationary Lorenz-gauge
potentials can be chosen as
\begin{equation}
 \bm A_j^{\mathrm p}=\frac{v}{c^2}V_j^{\mathrm p}\ez,
 \qquad V_j^{\mathrm p}=V_j^{\mathrm p}(\rp,\zeta).
 \label{eq:pipe-potentials}
\end{equation}
The Lorenz condition is satisfied because
$\partial_zV_j^{\mathrm p}=\partial_\zeta V_j^{\mathrm p}$ and
$\partial_tV_j^{\mathrm p}=-v\partial_\zeta V_j^{\mathrm p}$.  The scalar wave equation then
reduces to
$[\Deltap+(1-\beta^2)\partial_\zeta^2]V_j^{\mathrm p}=-\rho_j/\varepsilon_0$.
Hence, the nonradiating pure-pipe solution is
\begin{subequations}
\begin{align}
 \left(\Deltap+\gamma^{-2}\partial_\zeta^2\right)V_j^{\mathrm p}
 &=-\frac{\rho_j}{\varepsilon_0},
 \label{eq:pipe-potential}\\
 V_j^{\mathrm p}|_{\partial\Omega_j}&=0,
 \label{eq:pipe-boundary}\\
 \E_{\perp,j}^{\mathrm p}&=-\gradp V_j^{\mathrm p},
 &E_{z,j}^{\mathrm p}&=-\gamma^{-2}\partial_\zeta V_j^{\mathrm p},
 \label{eq:pipe-E}\\
 \B_{\perp,j}^{\mathrm p}&=\frac{v}{c^2}\ez\times\E_{\perp,j}^{\mathrm p},
 \label{eq:pipe-Bperp}\\
 B_{z,j}^{\mathrm p}&=0.
 \label{eq:pipe-B}
\end{align}
\end{subequations}
It follows that
\begin{equation}
 \bm F_{\!\perp,j}^{\mathrm p}=\gamma^{-2}\E_{\perp,j}^{\mathrm p},
 \qquad
 \overline E_{z,j}^{\mathrm p}(\rp,z,s)=E_{z,j}^{\mathrm p}(\rp,-s).
 \label{eq:pipe-force}
\end{equation}
Both space-charge forces are independent of $z$ on a witness trajectory, and their infinite
integrals diverge.  In the two uniform tails we therefore use the local scattered fields
\begin{equation}
 (\E_j^{\mathrm{sc}},\B_j^{\mathrm{sc}})
 =(\E,\B)
 -(\E_j^{\mathrm p},\B_j^{\mathrm p}),
 \qquad j=1,2.
 \label{eq:local-scattered}
\end{equation}
The device and pipe problems contain the same prescribed source.  Hence, the differences in
Eq.~\eqref{eq:local-scattered} satisfy the source-free Maxwell equations in the corresponding
uniform guides.  This subtraction is local to the two tails.  In the principal two-port convention,
no stationary field is subtracted from the finite direct interval, even when
$\Omega_1=\Omega_2$.  A global subtraction for identical pipes defines a different normalization,
and its exact relation to the two-port wake is given in Eqs.~\eqref{eq:equal-pipe-long-relation} and
\eqref{eq:equal-pipe-trans-relation} below.

\subsection{Source-free identities at finite velocity}

We first derive two identities which will be used for both indirect formulations.  In a uniform pipe, let
$(\E^{\mathrm{sc}},\B^{\mathrm{sc}})$ be any source-free field.  From Gauss' and
Amp\`ere--Maxwell laws,
\begin{align}
 \divp\Fp^{\mathrm{sc}}
 &=-\partial_zE_z^{\mathrm{sc}}
   -\frac{v}{c^2}\partial_tE_z^{\mathrm{sc}}.
 \label{eq:finitebeta-div-lab}
\end{align}
With Eq.~\eqref{eq:chain}, we obtain
\begin{equation}
 \divp\overline\Fp^{\mathrm{sc}}
 =-\frac{\dd\overline E_z^{\mathrm{sc}}}{\dd z}
   +\gamma^{-2}\partial_s\overline E_z^{\mathrm{sc}}.
 \label{eq:finitebeta-div}
\end{equation}
Faraday's law and magnetic-flux conservation give, independently,
\begin{align}
 \curlp\Fp^{\mathrm{sc}}
 &=-\partial_tB_z^{\mathrm{sc}}-v\partial_zB_z^{\mathrm{sc}},
 \nonumber\\
 \curlp\overline\Fp^{\mathrm{sc}}
 &=-v\frac{\dd\overline B_z^{\mathrm{sc}}}{\dd z}.
 \label{eq:finitebeta-curl}
\end{align}

The longitudinal component satisfies the homogeneous wave equation.  Since
\begin{align}
 \Deltap E_z^{\mathrm{sc}}
 &=-\partial_z^2E_z^{\mathrm{sc}}+c^{-2}\partial_t^2E_z^{\mathrm{sc}},
 \nonumber\\
 -\partial_z^2+v^{-2}\partial_t^2&=-D_+D_-,
 \label{eq:factorization}
\end{align}
its pullback obeys
\begin{equation}
 \left(\Deltap+\gamma^{-2}\partial_s^2\right)
 \overline E_z^{\mathrm{sc}}
 =-\frac{\dd}{\dd z}\overline{D_-E_z^{\mathrm{sc}}}.
 \label{eq:finitebeta-wave-pullback}
\end{equation}
In the following we use the operator
\begin{equation}
 \mathcal L_\beta=\Deltap+\gamma^{-2}\partial_s^2.
 \label{eq:Lbeta}
\end{equation}
For Dirichlet data on the pipe wall and decay as $|s|\to\infty$, this boundary-value problem has a
unique solution.  Indeed, if $\mathcal L_\beta h=0$, multiplication by $h$ and integration by parts gives
\begin{equation}
 0=-\int_{-\infty}^{\infty}\!\int_\Omega
 \left(|\gradp h|^2+\gamma^{-2}|\partial_sh|^2\right)\dd A\dd s,
 \label{eq:Lbeta-uniqueness}
\end{equation}
and hence $h=0$.

The one-sided fixed-time and fixed-plane formulations, including the optional globally
subtracted equal-pipe normalization, are derived in Appendix~\ref{app:one-sided}.  We proceed
directly to the principal two-port construction.

\subsection{Unified two-port finite-reference wake}
\label{sec:two-port}

We now use the fixed-plane form at both ends of the computational domain.  This is the principal
finite-$\beta$ algorithm for both equal and unequal pipes.  Each correction is calculated in its own
fixed port cross section; for identical pipes the same reference problem may of course be reused.  Put
$\sigma_1=-1$, $\sigma_2=+1$, $t_j=(z_j+s)/v$, and define
\begin{equation}
 w_1=\int_{-\infty}^{z_1}\overline E_{z,1}^{\mathrm{sc}}\dd z,
 \qquad
 w_2=\int_{z_2}^{\infty}\overline E_{z,2}^{\mathrm{sc}}\dd z.
 \label{eq:finitebeta-wtails}
\end{equation}
The fixed-plane notation of Appendix~\ref{app:fixed-plane} is recovered at the right port by setting
$z_0=z_2$: then $w=w_2$ and $\Kp^{\mathrm z}=\Kp^{(2)}$.
Integration of Eq.~\eqref{eq:finitebeta-wave-pullback} toward the corresponding infinity gives
\begin{equation}
 \mathcal L_\beta w_j
 =\sigma_jD_-E_{z,j}^{\mathrm{sc}}(\rp,z_j,t_j),
 \qquad w_j|_{\partial\Omega_j}=0.
 \label{eq:finitebeta-long-PDE}
\end{equation}
The opposite signs follow from the upper endpoint of the left tail and the lower endpoint of the
right tail.  For localized data, $w_j$ decays as $|s|\to\infty$.

The longitudinal wake is calculated with the complete device field on $[z_1,z_2]$, irrespective
of whether the two pipes are equal or unequal:
\begin{equation}
 QW_\parallel^{[z_1,z_2]}
 =-w_1-\int_{z_1}^{z_2}\overline E_z\dd z-w_2.
 \label{eq:finitebeta-long-total}
\end{equation}
Thus the central integral includes space charge.  A stationary field is subtracted only in the two
uniform tails: the field of pipe 1 on the left and the field of pipe 2 on the right.

We define the corresponding transverse tail impulses by
\begin{equation}
 \Kp^{(1)}=\int_{-\infty}^{z_1}\overline{\bm F}_{\!\perp,1}^{\mathrm{sc}}\dd z,
 \qquad
 \Kp^{(2)}=\int_{z_2}^{\infty}\overline{\bm F}_{\!\perp,2}^{\mathrm{sc}}\dd z.
 \label{eq:finitebeta-Ktails}
\end{equation}
Integration of Eqs.~\eqref{eq:finitebeta-div} and \eqref{eq:finitebeta-curl} gives
\begin{align}
 \divp\Kp^{(j)}
 &=\sigma_jE_{z,j}^{\mathrm{sc}}(\rp,z_j,t_j)
   +\gamma^{-2}\partial_s w_j,
 \label{eq:finitebeta-tail-div}\\
 \curlp\Kp^{(j)}
 &=\sigma_jvB_{z,j}^{\mathrm{sc}}(\rp,z_j,t_j).
 \label{eq:finitebeta-tail-curl}
\end{align}
The tail is reconstructed from
\begin{align}
 \Kp^{(j)}&=\gradp\Phi_j+\ez\times\gradp\Psi_j,
 \label{eq:finitebeta-Hodge}\\
 \Deltap\Phi_j
 &=\sigma_jE_{z,j}^{\mathrm{sc}}(\rp,z_j,t_j)
   +\gamma^{-2}\partial_s w_j,
 &\Phi_j|_{\partial\Omega_j}&=0,
 \label{eq:finitebeta-Phi}\\
 \Deltap\Psi_j
 &=\sigma_jvB_{z,j}^{\mathrm{sc}}(\rp,z_j,t_j),
 &\partial_n\Psi_j|_{\partial\Omega_j}&=0,
 \label{eq:finitebeta-Psi}\\
 &&\int_{\Omega_j}\Psi_j\dd A&=0.\nonumber
\end{align}
The finite-$\gamma$ term $\gamma^{-2}\partial_sw_j$ is required in the TM-like problem.  The
TE-like problem remains local in $s$ and cannot, in general, be obtained from the longitudinal tail.

The central impulse is likewise calculated with the complete device field for both equal and
unequal pipes:
\begin{equation}
 \Kp^{\mathrm{dir}}=\int_{z_1}^{z_2}
 \left(\overline\E_\perp
       +v\ez\times\overline\B_\perp\right)\dd z.
 \label{eq:finitebeta-direct}
\end{equation}
Finally, in the same convention as Eq.~\eqref{eq:finitebeta-long-total}, the transverse wake is
\begin{equation}
 Q\Wp^{[z_1,z_2]}
 =\Kp^{(1)}+\Kp^{\mathrm{dir}}+\Kp^{(2)}.
 \label{eq:finitebeta-trans-total}
\end{equation}

The same fixed-port two-port wake can be assembled with the moving-endpoint output tail
$(u,\Kp^{\mathrm t})$ defined in Appendix~\ref{app:fixed-time}.  Keep the
right reference plane $z_2$ fixed and choose the snapshot position $z_0$ sufficiently far downstream
that $a(s)=z_0-s\ge z_2$ remains in the uniform output pipe for every reconstructed lag.  Directly
from the tail definitions,
\begin{align}
 u
 &=w_2-\int_{z_2}^{a(s)}\overline E_{z,2}^{\mathrm{sc}}\dd z,
 \nonumber\\
 \Kp^{\mathrm t}
 &=\Kp^{(2)}-\int_{z_2}^{a(s)}\overline{\bm F}_{\!\perp,2}^{\mathrm{sc}}\dd z.
 \label{eq:moving-fixed-port-tail-relations}
\end{align}
The oriented stationary reference strip is
\begin{align}
 R_\parallel^{\mathrm t}
 &=\left[z_2-a(s)\right]E_{z,2}^{\mathrm p}(\rp,-s),
 \nonumber\\
 \bm R_{\!\perp}^{\mathrm t}
 &=\left[z_2-a(s)\right]\bm F_{\!\perp,2}^{\mathrm p}(\rp,-s).
 \label{eq:moving-reference-strip}
\end{align}
Here $R_\parallel^{\mathrm t}$ and $\bm R_{\!\perp}^{\mathrm t}$ are the oriented stationary output-pipe contributions from
$a(s)$ back to the fixed plane $z_2$; their common scalar factor $z_2-a(s)$ is negative when
$a(s)>z_2$.  The fixed-reference result is
\begin{align}
 QW_\parallel^{[z_1,z_2]}
 &=-w_1-\int_{z_1}^{a(s)}\overline E_z\dd z-u-R_\parallel^{\mathrm t},
 \label{eq:moving-two-port-long}\\
 Q\Wp^{[z_1,z_2]}
 &=\Kp^{(1)}+\int_{z_1}^{a(s)}\overline\Fp\dd z
   +\Kp^{\mathrm t}+\bm R_{\!\perp}^{\mathrm t}.
 \label{eq:moving-two-port-trans}
\end{align}
Equation~\eqref{eq:moving-fixed-port-tail-relations} subtracts the scattered field on
$[z_2,a(s)]$ from each fixed-plane tail.  The complete direct integral contains the same segment with
the complete device field, and the oriented reference strip cancels its stationary part.  Therefore
Eqs.~\eqref{eq:moving-two-port-long} and \eqref{eq:moving-two-port-trans} reduce exactly to
Eqs.~\eqref{eq:finitebeta-long-total} and \eqref{eq:finitebeta-trans-total}.  The reference strip is
essential at finite $\beta$; without it the moving-endpoint formula would use the lag-dependent right
reference plane $a(s)$ rather than the prescribed fixed plane $z_2$.
The problems for $j=1$ and $j=2$ are solved independently on $\Omega_1$ and $\Omega_2$.  We do not
continue one pipe reference through the structure and do not search for an interface where the two
reference fields should be joined.

For identical pipes, let
$\E_1^{\mathrm p}=\E_2^{\mathrm p}=\E^{\mathrm p}$,
$\B_1^{\mathrm p}=\B_2^{\mathrm p}=\B^{\mathrm p}$, and $L=z_2-z_1$.  A global scattered field
can then be defined.  Splitting the complete field on the central interval as
$(\E,\B)=(\E^{\mathrm{sc}},\B^{\mathrm{sc}})
 +(\E^{\mathrm p},\B^{\mathrm p})$ gives
\begin{align}
 QW_\parallel^{[z_1,z_2]}
 &=QW_\parallel^{\mathrm{sc}}
   -L E_z^{\mathrm p}(\rp,-s),
 \label{eq:equal-pipe-long-relation}\\
 Q\Wp^{[z_1,z_2]}
 &=Q\Wp^{\mathrm{sc}}
   +L\Fp^{\mathrm p}(\rp,-s).
 \label{eq:equal-pipe-trans-relation}
\end{align}
Thus the two-port algorithm has a regular equal-pipe limit, but its normalization is not identical
to that of the globally subtracted wake.  In particular, the complete field, not the scattered
field, must be integrated on $[z_1,z_2]$ when the two-port convention is used.  If the conventional
port-independent equal-pipe wake is desired, it is recovered by adding
$L E_z^{\mathrm p}/Q$ to $W_\parallel^{[z_1,z_2]}$ and subtracting
$L\Fp^{\mathrm p}/Q$ from $\Wp^{[z_1,z_2]}$.  The stationary pipe field itself satisfies
$\partial_s\Fp^{\mathrm p}=-\gradp E_z^{\mathrm p}$ on the sampled trajectory, so this
conversion preserves the endpoint-free Panofsky--Wenzel relation for identical pipes.

At $\beta<1$, moving either port plane changes the finite-reference result by the stationary field
contained in the transferred pipe segment.  For shifts that remain inside the corresponding uniform
pipes,
\begin{align}
 \delta\!\left(QW_\parallel^{[z_1,z_2]}\right)
 &=E_{z,1}^{\mathrm p}(\rp,-s)\,\delta z_1
 \nonumber\\
 &\quad-E_{z,2}^{\mathrm p}(\rp,-s)\,\delta z_2,
 \label{eq:port-shift-long}\\
 \delta\!\left(Q\Wp^{[z_1,z_2]}\right)
 &=-\bm F_{\!\perp,1}^{\mathrm p}(\rp,-s)\,\delta z_1
 \nonumber\\
 &\quad+\bm F_{\!\perp,2}^{\mathrm p}(\rp,-s)\,\delta z_2.
 \label{eq:port-shift-trans}
\end{align}
Here positive $\delta z_j$ denotes a shift toward increasing $z$.  This is a consequence of the
finite-reference definition, not an error of the indirect calculation.
Therefore, the port planes must be specified with a finite-$\beta$ result.  In the equal-pipe case,
Eqs.~\eqref{eq:equal-pipe-long-relation} and \eqref{eq:equal-pipe-trans-relation} remove this
reference-length dependence and recover the port-independent scattered wake.  In the limit
$\beta\to1$, the stationary longitudinal field and transverse Lorentz force vanish, the $s$
coupling disappears,
$\mathcal L_\beta\to\Deltap$, and Eqs.~\eqref{eq:finitebeta-long-PDE}--
\eqref{eq:finitebeta-Psi} reduce to the two-port Poisson formulas of
Secs.~\ref{sec:known-beta1} and \ref{sec:transverse-beta1}.

\subsection{Numerical solution}

Only the longitudinal equation couples different witness lags.  Each
$h\in\{u,w,w_1,w_2\}$ satisfies $\mathcal L_\beta h=f_h$ with the source appropriate to its
representation and port.  With the Fourier convention
\begin{equation}
 \widehat h(\rp,k)=\int_{-\infty}^{\infty}h(\rp,s)e^{-\ii ks}\dd s,
 \label{eq:Fourier-convention}
\end{equation}
the transformed equation is
\begin{equation}
 \left(-\Deltap+\frac{k^2}{\gamma^2}\right)\widehat h(\rp,k)
 =-\widehat f_h(\rp,k),
 \qquad \widehat h|_{\partial\Omega}=0.
 \label{eq:modified-Helmholtz}
\end{equation}
Thus each Fourier harmonic requires a positive-definite modified Helmholtz solve in a two-dimensional
cross section; the zero harmonic gives the usual Dirichlet Poisson problem.  After $h$ and
$\partial_sh$ have been reconstructed, the transverse TM and TE potentials follow from two
independent two-dimensional Poisson problems at each $s$.

The computational sequence for equal or unequal pipes is as follows.  First, the stationary
fields of the two pipes are calculated from Eqs.~\eqref{eq:pipe-potential}--\eqref{eq:pipe-B}.
The complete device fields, including space charge, are then integrated from $z_1$ to $z_2$.
At each port the stationary field of the matching pipe is subtracted from the recorded data.
Equation~\eqref{eq:finitebeta-long-PDE} is solved directly in $(x,y,s)$ or through
Eq.~\eqref{eq:modified-Helmholtz}; finally, Eqs.~\eqref{eq:finitebeta-Phi} and
\eqref{eq:finitebeta-Psi} are solved and the wakes are assembled from
Eqs.~\eqref{eq:finitebeta-long-total} and \eqref{eq:finitebeta-trans-total}.

\section{Panofsky--Wenzel theorem for the two-port convention}
\label{sec:PW}

We first recall the local identity which is valid for every constant
$v$~\cite{PanofskyWenzel1956,VaganianHenke1995}:
\begin{equation}
 \partial_s\overline\Fp
 =\frac{\dd\overline\E_\perp}{\dd z}-\gradp\overline E_z.
 \label{eq:local-PW}
\end{equation}
For example, its $x$ component follows from
\begin{align}
 \partial_s(E_x-vB_y)
 &=v^{-1}\partial_tE_x-\partial_tB_y\nonumber\\
 &=v^{-1}\partial_tE_x+\partial_zE_x-\partial_xE_z,
 \label{eq:PW-x-derivation}
\end{align}
where Faraday's law was used in the second line.  The $y$ component follows in the same way.
Thus Eq.~\eqref{eq:local-PW} does not require $\beta=1$.

The corresponding consistency check for the fixed-time moving-endpoint and fixed-plane one-sided
representations, including the cancellation of their endpoint terms, is given in
Appendix~\ref{app:PW-one-sided}.  We now apply Eq.~\eqref{eq:local-PW} directly to the principal
two-port convention.  For any fixed finite interval $[z_A,z_B]$,
\begin{equation}
 \begin{aligned}
 \partial_s\left(\frac{1}{Q}\int_{z_A}^{z_B}\overline\Fp\dd z\right)
 ={}&\gradp\left(-\frac{1}{Q}\int_{z_A}^{z_B}\overline E_z\dd z\right)\\
 &+\frac{\overline\E_\perp(\rp,z_B,s)-\overline\E_\perp(\rp,z_A,s)}{Q}.
 \end{aligned}
 \label{eq:finite-PW}
\end{equation}
Application of Eq.~\eqref{eq:local-PW} to the three intervals of the two-port construction gives
\begin{align}
 \partial_s\Kp^{(1)}
 &=\E_{\perp,1}^{\mathrm{sc}}(\rp,z_1,t_1)\nonumber\\
 &\quad-\gradp w_1,
 \label{eq:PW-left}\\
 \partial_s\Kp^{\mathrm{dir}}
 &=\E_\perp(\rp,z_2,t_2)
   -\E_\perp(\rp,z_1,t_1)\nonumber\\
 &\quad-\gradp\int_{z_1}^{z_2}\overline E_z\dd z,
 \label{eq:PW-center}\\
 \partial_s\Kp^{(2)}
 &=-\E_{\perp,2}^{\mathrm{sc}}(\rp,z_2,t_2)\nonumber\\
 &\quad-\gradp w_2.
 \label{eq:PW-right}
\end{align}
We now use $\E_j^{\mathrm{sc}}=\E-\E_j^{\mathrm p}$ at each port.  The device fields
at both port planes cancel.  Dividing the sum by $Q$ and using
Eq.~\eqref{eq:finitebeta-long-total}, we obtain
\begin{equation}
 \begin{aligned}
 \partial_s\Wp^{[z_1,z_2]}
 ={}&\gradp W_\parallel^{[z_1,z_2]}\\
 &+\frac{\E_{\perp,2}^{\mathrm p}(\rp,-s)
             -\E_{\perp,1}^{\mathrm p}(\rp,-s)}{Q}.
 \end{aligned}
 \label{eq:unequal-PW}
\end{equation}
Equation~\eqref{eq:unequal-PW} is the longitudinal--transverse wake-potential relation for the
present two-port convention with equal or unequal asymptotic pipes.  The explicit boundary term
vanishes for identical pipe fields.  At $\beta=1$ the
stationary transverse Lorentz force in either pipe is zero, but the stationary electric fields can
still differ; hence the boundary term in Eq.~\eqref{eq:unequal-PW} remains necessary for unequal
pipes~\cite{VaganianHenke1995,HenkeBruns2006}.

\section{Numerical examples: transverse wake of a rectangular step-out}
\label{sec:examples}

In this section we test the reconstructed transverse wake.  We use the complete transverse
domain without electric or magnetic symmetry planes.  The example also illustrates numerically the
unequal-pipe boundary term in Eq.~\eqref{eq:unequal-PW}.

\subsection{Geometry and numerical setup}

We consider an abrupt rectangular step at $z=0$.  In the transverse coordinates of
Sec.~\ref{sec:definitions}, the two apertures are
\begin{equation}
 \begin{split}
 \Omega_1&=(-50,50)\,\mathrm{mm}\times(-10,10)\,\mathrm{mm},\\
 \Omega_2&=(-50,50)\,\mathrm{mm}\times(-50,50)\,\mathrm{mm}.
 \end{split}
 \label{eq:example-apertures}
\end{equation}
Thus only the vertical aperture changes, but the port cross sections are unequal.  Both the source
and the witness are at
\begin{equation}
 \bm r_q=\bm r_w=(0,6\,\mathrm{mm}).
 \label{eq:example-trajectories}
\end{equation}
This relatively large offset makes the transverse signal and the unequal-pipe boundary contribution clearly
visible and leaves a $4\,\mathrm{mm}$ distance to the narrow-pipe wall.  The bunch  longitudinal distribution is Gaussian with
$\sigma_z=2\,\mathrm{mm}$.

\begin{figure*}[t]
 \centering
 \includegraphics[width=0.98\textwidth]{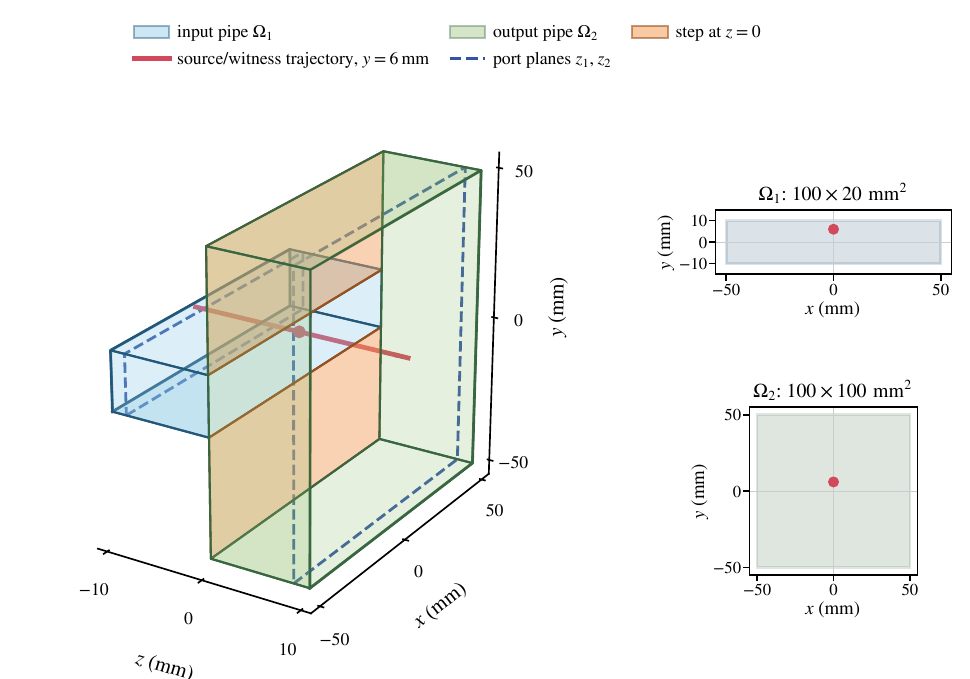}
 \caption{Three-dimensional vacuum geometry and the two port cross sections.  The coordinates are
 those of the paper: $z$ is longitudinal and $x,y$ are transverse.  The red line is the common
 source and witness trajectory.  Dashed blue rectangles mark the port planes of the longest
 fixed-domain calculation, $z_1=-8.538\,\mathrm{mm}$ and
 $z_2=8.409\,\mathrm{mm}$.}
 \label{fig:example-geometry}
\end{figure*}

The $\beta=1$ time-domain calculations use the TE/TM field solver of Ref.~\cite{ZagorodnovWeiland2005}.
The nominal mesh spacings are $\Delta z=0.269\,\mathrm{mm}$ and
$\Delta x=\Delta y=0.4\,\mathrm{mm}$.  The reported interval contains 78 samples at
$s_n=-5\sigma_z+n\Delta z$.  The fixed-domain calculation uses the TE/TM
update, open longitudinal boundaries, and the two fixed-plane tail reconstructions of
Eqs.~\eqref{eq:beta1-Phi}--\eqref{eq:beta1-Psi}.  The complete fields are integrated directly between
the two port planes, while the stationary field of the matching pipe is subtracted only in each
tail.  On the selected trajectory,
\begin{align}
 W_y^{\mathrm{dir}}(s)
 &=\frac{1}{Q}\int_{z_1}^{z_2}
   \left(E_y+cB_x\right)
   \left(\bm r_w,z,\frac{z+s}{c}\right)\dd z,
 \label{eq:example-direct-y}\\
 W_y^{\mathrm{tail}}(s)
 &=\frac{K_y^{(1)}(s)+K_y^{(2)}(s)}{Q},
 \nonumber\\
 W_y^{[z_1,z_2]}(s)
 &=W_y^{\mathrm{dir}}(s)+W_y^{\mathrm{tail}}(s).
 \label{eq:example-two-port-y}
\end{align}
The relative tolerance for every port Poisson problem is $10^{-10}$.  The largest residual in the
nominal calculations is $9.87\times10^{-11}$.  The device and stationary-pipe calculations use the
same charge deposition, transverse grids, time staggering, and interpolation at the port planes.
We also monitor the Neumann compatibility integral
$\int_{\Omega_j}B_{z,j}^{\mathrm{sc}}\dd A$ and the representation identities
\eqref{eq:u-w-relation} and \eqref{eq:Kt-Kz-relation}.

\begin{figure*}[t]
 \centering
 \includegraphics[width=0.98\textwidth]{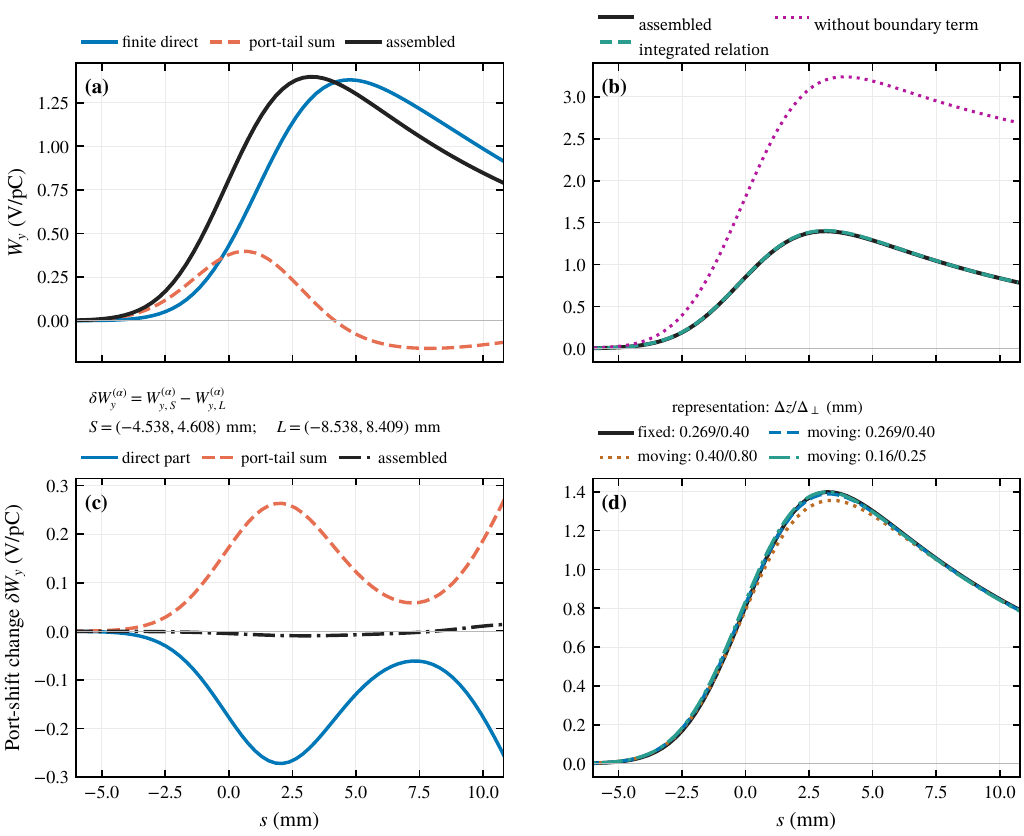}
 \caption{Transverse-wake validation for the full-domain rectangular step-out.  (a) Fixed-plane
 direct integral, sum of the two reconstructed tails, and assembled wake.  (b) Assembled transverse wake
 compared with integrations of the longitudinal--transverse relation with and without the
 unequal-pipe boundary term.
 (c) Port-plane invariance test using the differences defined in
 Eqs.~\eqref{eq:example-port-difference} and \eqref{eq:example-port-difference-assembled}.  Superscripts
 $S$ and $L$ denote the shortest and longest fixed domains, respectively.  The opposite changes of the direct part and
 port-tail sum cancel in the assembled wake.  (d) Fixed-plane result and moving-endpoint results on
 three grids.  In the mesh labels, the first number is $\Delta z$ and the second is the common
 transverse spacing.}
 \label{fig:example-validation}
\end{figure*}

Figure~\ref{fig:example-validation}(a) shows the directly accumulated contribution, the sum of the
two tails, and the complete result.  Their maximum absolute values are $1.3810$, $0.3970$, and
$1.3984\,\mathrm{V/pC}$, respectively.  Thus the tail gives an important contribution, although the
maximum of the complete wake is close to the maximum of the direct curve.

\subsection{Panofsky--Wenzel test and the boundary term}

For an independent calculation, the two-port longitudinal wake is sampled at
$y_w=5.6$ and $6.4\,\mathrm{mm}$, with the source position kept fixed.  A centered difference with
$\delta y=0.4\,\mathrm{mm}$ is inserted into the $y$ component of
Eq.~\eqref{eq:unequal-PW}.  Integration from the causal head gives
\begin{equation}
 \begin{aligned}
 W_y^{[z_1,z_2]}(s)={}&\int_{-\infty}^{s}\!\Bigg[
 \frac{W_\parallel^{[z_1,z_2]}(0,y_w+\delta y,s')}{2\delta y}\\
 &-\frac{W_\parallel^{[z_1,z_2]}(0,y_w-\delta y,s')}{2\delta y}\\
 &+\frac{E_{y,2}^{\mathrm p}(\bm r_w,-s')
              -E_{y,1}^{\mathrm p}(\bm r_w,-s')}{Q}\Bigg]\dd s'.
 \end{aligned}
 \label{eq:example-unequal-relation}
\end{equation}
The stationary fields in the second line are calculated separately in the two rectangular
cross sections with the same five-point Dirichlet Poisson operator used for the reference-pipe
fields.  For the normalized Gaussian bunch, the separate integrals of the input- and output-pipe
terms are $1.9459$ and $0.038003\,\mathrm{V/pC}$.  Hence, their difference and the integrated
boundary contribution are $-1.9079\,\mathrm{V/pC}$.

Let $f=W_y^{[z_1,z_2]}$ and let $g$ denote the transverse wake obtained by integrating the
right-hand side of Eq.~\eqref{eq:unequal-PW}.  We use
$\epsilon_2=\|f-g\|_2/\|f\|_2$ and
$\epsilon_\infty=\max|f-g|/\max|f|$.  The latter transverse wake agrees with the directly reconstructed
wake to
$\epsilon_2=3.78\times10^{-3}$ and $\epsilon_\infty=5.33\times10^{-3}$; see
Fig.~\ref{fig:example-validation}(b).  If the boundary term is omitted, the corresponding errors are
$1.538$ and $1.366$.  The resulting curve reaches $3.240\,\mathrm{V/pC}$, more than twice the
physical peak of $1.398\,\mathrm{V/pC}$.  This comparison shows that the boundary term in
Eq.~\eqref{eq:unequal-PW} is also necessary at $\beta=1$.  The pure-pipe transverse Lorentz force
vanishes in this limit, but the stationary transverse electric fields of unequal pipes are
different.

\subsection{Port-plane, representation, and grid checks}

We repeated the fixed-domain calculation for three pairs of port planes in the uniform pipe
sections.  Table~\ref{tab:example-ports} lists the maximum absolute values.  The direct and tail
terms change strongly and in opposite directions when the planes are moved.  To show this
cancellation point by point, in Fig.~\ref{fig:example-validation}(c) we use
\begin{equation}
 \begin{split}
 \delta W_y^{(\alpha)}(s)={}&
 W_y^{(\alpha)}(s;z_1^S,z_2^S)\\
 &-W_y^{(\alpha)}(s;z_1^L,z_2^L),\\
 &\alpha\in\{\mathrm{dir},\mathrm{tail}\}.
 \end{split}
 \label{eq:example-port-difference}
\end{equation}
For the assembled wake we analogously define
\begin{equation}
 \delta W_y(s)=
 W_y^{[z_1^S,z_2^S]}(s)-W_y^{[z_1^L,z_2^L]}(s).
 \label{eq:example-port-difference-assembled}
\end{equation}
Here $(z_1^S,z_2^S)=(-4.538,4.608)\,\mathrm{mm}$ and
$(z_1^L,z_2^L)=(-8.538,8.409)\,\mathrm{mm}$.  Thus the blue curve is negative because shortening
the direct integration interval removes part of the direct kick.  The port-tail correction then
increases by nearly the same amount, giving the positive orange curve.  Since
$\delta W_y=\delta W_y^{\mathrm{dir}}+\delta W_y^{\mathrm{tail}}$, the black dash-dotted curve is the
pointwise sum of the blue and orange curves and stays close to zero.  The maximum absolute changes
of the direct and tail parts are $0.2724$ and $0.2868\,\mathrm{V/pC}$, whereas the complete wake
changes by only $0.01440\,\mathrm{V/pC}$.  The ratio of the remaining change to the larger component
change is $5.02\%$, and the remaining change is $1.03\%$ of the wake maximum.  Thus the assembled
result is nearly independent of the port planes, while the direct and tail parts separately
are not.

\begin{table}[b]
 \caption{Port-plane test.  All wake entries are maximum absolute values in V/pC over the displayed
 interval.}
 \label{tab:example-ports}
 \begin{ruledtabular}
 \begin{tabular}{ccccc}
 $z_1$ (mm) & $z_2$ (mm) & $W_y^{\mathrm{dir}}$ & $W_y^{\mathrm{tail}}$ & $W_y^{[z_1,z_2]}$\\
 \hline
 $-4.538$ & $4.608$ & $1.2584$ & $0.6282$ & $1.3891$\\
 $-6.538$ & $6.643$ & $1.3416$ & $0.4892$ & $1.4026$\\
 $-8.538$ & $8.409$ & $1.3810$ & $0.3970$ & $1.3984$
 \end{tabular}
 \end{ruledtabular}
\end{table}

We also repeated the calculation with the fixed-time moving-endpoint algorithm of
Appendix~\ref{app:fixed-time}.  The moving window uses the TE/TM update of
Ref.~\cite{ZagorodnovWeiland2005} and the same nominal transverse mesh.  Its assembled wake and the
fixed-plane result agree to
$8.35\times10^{-3}$ in relative $L_2$ norm and $1.22\times10^{-2}$ in relative maximum norm.  This
result confirms numerically the equivalence expressed by Eqs.~\eqref{eq:u-w-relation} and
\eqref{eq:Kt-Kz-relation}, with each direct integral paired with its own tail representation.

Finally, the moving-window calculation was performed on coarse, nominal, and fine meshes with
$(\Delta z,\Delta x,\Delta y)=(0.40,0.80,0.80)$,
$(0.269,0.40,0.40)$, and $(0.16,0.25,0.25)\,\mathrm{mm}$, respectively.  From coarse to nominal,
the relative $L_2$ and maximum changes of the assembled wake are $2.15\%$ and $3.06\%$; from
nominal to fine they decrease to $0.852\%$ and $1.29\%$.  Figure~\ref{fig:example-validation}(d)
shows both grid convergence and agreement of the fixed-plane and moving-endpoint representations.
The symmetry-forbidden component $W_x$ in the nominal fixed-plane calculation is
$1.22\times10^{-10}$ of the $W_y$ maximum.

\subsection{Finite-velocity two-port calculation}

To test the finite-velocity algorithm without changing the benchmark geometry, we repeat the
nominal fixed-plane calculation with $\beta=0.8$ ($\gamma=1.6667$).  The apertures, source and
witness positions, $Q$, $\sigma_z$, port planes, and the 78 reported wake samples are the same as
above.  The fixed integration planes are $z_1=-8.538\,\mathrm{mm}$ and
$z_2=8.409\,\mathrm{mm}$.  The finite-velocity fields are advanced in a fixed computational domain
with the explicit Yee FDTD scheme~\cite{Yee1966}.  The spatial mesh is
$(\Delta z,\Delta x,\Delta y)=(0.269,0.40,0.40)\,\mathrm{mm}$, and the time step is
$c\Delta t=0.125\,\mathrm{mm}$.  The same transverse mesh is used for the independently calculated
stationary fields in the two pipes.  The complete Lorentz force is integrated between the fixed
planes, including the space-charge field.  At each plane the stationary field of the matching pipe
is subtracted before the longitudinal modified Helmholtz and transverse Hodge problems are solved.
Thus this calculation includes both the lag coupling in
$\mathcal L_\beta$ and the $\gamma^{-2}\partial_s w_j$ source in Eq.~\eqref{eq:finitebeta-Phi}.
The lag histories are zero padded before the Fourier transform, and each harmonic of
Eq.~\eqref{eq:modified-Helmholtz} is solved in its corresponding port cross section.

\begin{figure*}[t]
 \centering
 \includegraphics[width=0.98\textwidth]{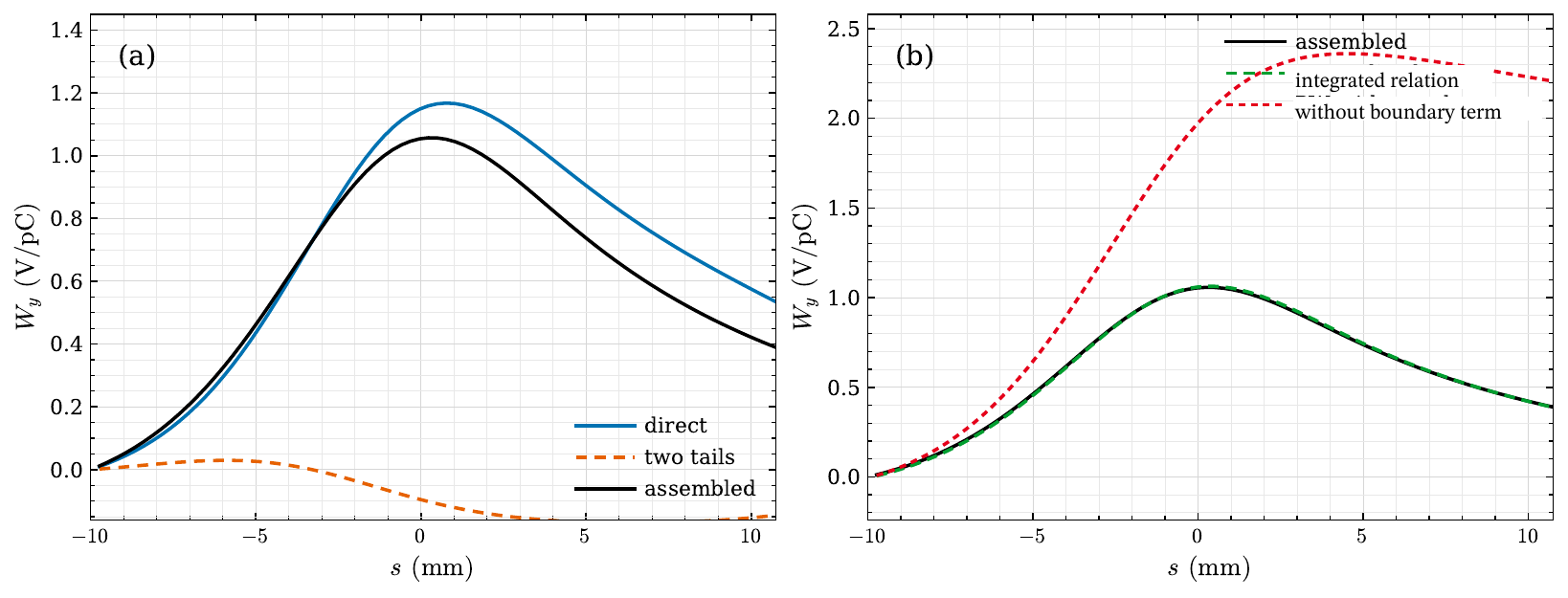}
 \caption{Finite-velocity calculation for the same unequal rectangular step-out as in
 Fig.~\ref{fig:example-geometry}, with $\beta=0.8$.  (a) Complete-field direct integral, sum of the
 two local-reference tails, and assembled two-port wake.  (b) Assembled result compared with the
 independently integrated longitudinal--transverse relation and with the expression that omits its
 boundary term.}
 \label{fig:example-finite-beta}
\end{figure*}

Figure~\ref{fig:example-finite-beta}(a) shows the complete-field direct result, the sum of the two reconstructed
tails, and the assembled transverse wake.  Their maximum absolute values are $1.1673$, $0.1695$,
and $1.0569\,\mathrm{V/pC}$, respectively.  The $L_2$ norm of the tail correction is $14.6\%$ of
the assembled-wake norm, so the tail contribution is numerically significant.

For Fig.~\ref{fig:example-finite-beta}(b), the longitudinal wake is differenced at
$y_w\pm0.4\,\mathrm{mm}$ and integrated with the independently reconstructed difference
$E_{y,2}^{\mathrm p}-E_{y,1}^{\mathrm p}$, exactly as in
Eq.~\eqref{eq:example-unequal-relation}.  After removal of the common causal-head constant, the
assembled wake and the independently integrated right-hand side of Eq.~\eqref{eq:unequal-PW}
agree with relative $L_2$ and maximum errors of $0.879\%$ and $0.885\%$ when
normalized by the assembled result.  The ordinary longitudinal-gradient integral and the
unequal-pipe boundary primitive reach $2.3608$ and $1.8191\,\mathrm{V/pC}$, respectively, and
partly cancel; this makes the comparison particularly sensitive to the sign of the
$\gamma^{-2}\partial_s w_j$ term.  If the unequal-pipe boundary term is omitted, the errors become
$1.690$ and $1.720$. 

\section{Conclusions}

In this paper we have extended the indirect integration method to a transverse Lorentz-force
integral which has already been accumulated directly.  We started from the known longitudinal
moving-window and fixed-plane formulations.  In contrast to the earlier methods, we reconstruct
only the missing semi-infinite transverse tails and do not obtain the complete transverse wake from
the Panofsky--Wenzel theorem.

At $\beta=1$, the divergence and curl of each transverse tail are determined by the complete
longitudinal field components $E_{z,j}$ and $B_{z,j}$ at the corresponding port.  With a two-dimensional Hodge
decomposition we obtain a Dirichlet Poisson problem for the TM part and a Neumann Poisson problem
for the TE part.  The method is valid for different input and output cross sections and has both a
fixed-time moving-endpoint and a fixed-plane time-history form.  The TE correction cannot be omitted
if the direct force integral is stopped before the TE contribution is completely cancelled.

For a prescribed rigid bunch with $0<\beta<1$, we use the stationary bunch field in each uniform
pipe as the local reference.  The longitudinal source is $\beta^2D_-E_z^{\mathrm{sc}}$ for the moving endpoint and
$D_-E_z^{\mathrm{sc}}$ for the fixed plane.  The transverse TM problem contains the additional derivative
$\gamma^{-2}\partial_s$ of the longitudinal tail, while the TE problem is driven by $vB_z^{\mathrm{sc}}$.  In the
unified two-port convention, the complete space-charge field is kept in the finite direct interval
$[z_1,z_2]$ for both equal and unequal pipes, and the corresponding stationary pipe field is
subtracted only in the two reconstructed tails.  In this way we obtain a two-port finite-reference wake
without an arbitrary switching interface inside the structure.  For identical pipes, the
port-independent globally subtracted wake is obtained by removing the known stationary pure-pipe
contribution over $z_2-z_1$.  For unequal pipes the longitudinal--transverse relation contains the difference
of the stationary transverse electric fields at the two ports.

Finally, we have tested the transverse method for an unequal rectangular step-out at $\beta=1$ and
$\beta=0.8$, and used an additional unequal-pipe regression at the same finite velocity.  At finite velocity the
assembled two-port wake includes the coupled stationary pipe fields, the two independent modified
Helmholtz terminal solves, and the derivative contribution to the transverse TM source.  The
assembled results satisfy the two-port longitudinal--transverse relation with the unequal-pipe boundary term, whereas omitting
the stationary-field boundary term gives an order-unity error.  These tests provide a numerical
validation of the finite-$\beta$ two-port algorithm as well as of its $\beta=1$ limit.

\appendix

\section{One-sided finite-velocity representations}
\label{app:one-sided}

This appendix collects the one-sided fixed-time and fixed-plane formulations and their
conversion identities.  They are useful both as alternative implementations and as checks of
the principal two-port construction.

\subsection{Fixed time and moving endpoint}
\label{app:fixed-time}

Superscripts $\mathrm t$ and $\mathrm z$ on transverse impulses and Hodge potentials label the
fixed-time and fixed-plane representations, respectively; the corresponding longitudinal tails are
denoted simply by $u$ and $w$.
We first derive a one-sided output-tail reconstruction.  The tail equations are local and do not
depend on whether the final wake is normalized by a global subtraction or by the two-port
finite-reference convention.  Consider one output-pipe snapshot $t_0=z_0/v$ and put
\begin{equation}
 a(s)=z_0-s.
 \label{eq:moving-a}
\end{equation}
The point $a(s)$ must remain in the uniform output pipe for all lags being reconstructed.  Define
the downstream longitudinal tail by
\begin{equation}
 u(\rp,s)=\int_{a(s)}^\infty
 \overline E_z^{\mathrm{sc}}(\rp,z,s)\dd z.
 \label{eq:finitebeta-moving-u}
\end{equation}
The integral follows the witness characteristic, but its lower endpoint is always sampled at the
same time $t_0$, because $t[a(s),s]=t_0$.  From Leibniz' rule we obtain
\begin{align}
 \partial_su
 &=E_z^{\mathrm{sc}}(\rp,a,t_0)
   +\frac{1}{v}\int_a^\infty\partial_tE_z^{\mathrm{sc}}\dd z,
 \label{eq:moving-us}\\
 \partial_s^2u
 &=-\partial_zE_z^{\mathrm{sc}}(\rp,a,t_0)
   +\frac{1}{v}\partial_tE_z^{\mathrm{sc}}(\rp,a,t_0)
 \nonumber\\
 &\quad+\frac{1}{v^2}\int_a^\infty\partial_t^2E_z^{\mathrm{sc}}\dd z.
 \label{eq:moving-uss}
\end{align}
We now apply the wave equation under the integral and use Eq.~\eqref{eq:moving-uss}.  After
collecting the endpoint terms, we obtain
\begin{equation}
 \mathcal L_\beta u
 =\beta^2D_-E_z^{\mathrm{sc}}(\rp,a,t_0),
 \qquad u|_{\partial\Omega_2}=0.
 \label{eq:finitebeta-moving-PDE}
\end{equation}
The factor $\beta^2$ appears because the moving lower limit is differentiated.  For $\beta=1$,
Eq.~\eqref{eq:finitebeta-moving-PDE} reduces to the moving-window result \eqref{eq:Z-moving}.

The corresponding transverse tail is
\begin{equation}
 \Kp^{\mathrm t}(\rp,s)=\int_{a(s)}^\infty
 \overline\Fp^{\mathrm{sc}}(\rp,z,s)\dd z.
 \label{eq:finitebeta-moving-K}
\end{equation}
Integrating Eqs.~\eqref{eq:finitebeta-div} and \eqref{eq:finitebeta-curl}, while using
$\partial_su=E_z^{\mathrm{sc}}(\rp,a,t_0)+\int_a^\infty
\partial_s\overline E_z^{\mathrm{sc}}\dd z$, yields
\begin{align}
 \divp\Kp^{\mathrm t}
 &=\beta^2E_z^{\mathrm{sc}}(\rp,a,t_0)
   +\gamma^{-2}\partial_su,
 \label{eq:moving-K-div}\\
 \curlp\Kp^{\mathrm t}
 &=vB_z^{\mathrm{sc}}(\rp,a,t_0).
 \label{eq:moving-K-curl}
\end{align}
Thus
\begin{align}
 \Kp^{\mathrm t}&=\gradp\Phi^{\mathrm t}+\ez\times\gradp\Psi^{\mathrm t},
 \label{eq:moving-Hodge}\\
 \Deltap\Phi^{\mathrm t}
 &=\beta^2E_z^{\mathrm{sc}}(\rp,a,t_0)
   +\gamma^{-2}\partial_su,
 &\Phi^{\mathrm t}|_{\partial\Omega_2}&=0,
 \label{eq:moving-Phi}\\
 \Deltap\Psi^{\mathrm t}&=vB_z^{\mathrm{sc}}(\rp,a,t_0),
 &\partial_n\Psi^{\mathrm t}|_{\partial\Omega_2}&=0,
 \label{eq:moving-Psi}\\
 &&\int_{\Omega_2}\Psi^{\mathrm t}\dd A&=0.\nonumber
\end{align}
If identical asymptotic pipes permit a globally defined stationary reference, this tail may be
combined with a one-sided scattered-field accumulation.  We then introduce
\begin{align}
 u^{\mathrm{dir}}&=\int_{-\infty}^{a(s)}\overline E_z^{\mathrm{sc}}\dd z,
 &\Kp^{\mathrm{dir},\mathrm t}&=\int_{-\infty}^{a(s)}\overline\Fp^{\mathrm{sc}}\dd z.
 \label{eq:moving-direct-defs}
\end{align}
The globally subtracted scattered-field integrals can then be written as
\begin{equation}
 QW_\parallel^{\mathrm{sc}}=-u^{\mathrm{dir}}-u,
 \qquad Q\Wp^{\mathrm{sc}}=\Kp^{\mathrm{dir},\mathrm t}+\Kp^{\mathrm t}.
 \label{eq:moving-complete}
\end{equation}
Equations~\eqref{eq:moving-direct-defs} and \eqref{eq:moving-complete} define the optional
globally subtracted wake $W^{\mathrm{sc}}$; they are not the two-port finite-reference definition.
In the latter definition the direct field on the finite interval is the complete device
field, as specified in Sec.~\ref{sec:two-port}.  All sources in
Eqs.~\eqref{eq:finitebeta-moving-PDE}, \eqref{eq:moving-Phi}, and \eqref{eq:moving-Psi} are obtained
from the same snapshot $t_0$.  A scattered-field direct accumulation used with the one-sided tails
must terminate at the same moving point $a(s)$.  Stopping it instead at $z_0$ double counts the
segment $[a(s),z_0]$.

The time derivative needed in $D_-E_z$ need not be stored separately.  In the source-free output
pipe, the longitudinal Amp\`ere--Maxwell equation gives
\begin{equation}
 D_-E_z^{\mathrm{sc}}
 =\partial_zE_z^{\mathrm{sc}}
 -\frac{c^2}{v}\left(\partial_xB_y^{\mathrm{sc}}-\partial_yB_x^{\mathrm{sc}}\right).
 \label{eq:Dminus-snapshot}
\end{equation}
Hence, the complete right-hand side of the fixed-time problem can be calculated from one
synchronized snapshot of the electric and magnetic fields.

\subsection{Fixed plane and time history}
\label{app:fixed-plane}

Alternatively, we keep the output plane $z=z_0$ fixed and record its time history at
\begin{equation}
 t_s=(z_0+s)/v.
 \label{eq:fixed-plane-time}
\end{equation}
The longitudinal and transverse tails are now
\begin{align}
 w(\rp,s)&=\int_{z_0}^{\infty}
       \overline E_z^{\mathrm{sc}}(\rp,z,s)\dd z,
 \label{eq:fixed-plane-w}\\
 \Kp^{\mathrm z}(\rp,s)&=\int_{z_0}^{\infty}
       \overline\Fp^{\mathrm{sc}}(\rp,z,s)\dd z.
 \label{eq:fixed-plane-K}
\end{align}
Here the limits do not depend on $s$.  Integrating Eq.~\eqref{eq:finitebeta-wave-pullback} and using
the outgoing condition at infinity, we obtain
\begin{equation}
 \mathcal L_\beta w
 =D_-E_z^{\mathrm{sc}}(\rp,z_0,t_s),
 \qquad w|_{\partial\Omega_2}=0.
 \label{eq:fixed-plane-long-PDE}
\end{equation}
There is no factor $\beta^2$, since no limit is differentiated.  At $\beta=1$ this is the
fixed-plane Poisson equation [Eq.~(22) of Ref.~\cite{Zagorodnov2006}] and is equivalent to the
output-port part of the Henke--Bruns construction.

Integrating the transverse identities gives
\begin{align}
 \divp\Kp^{\mathrm z}
 &=E_z^{\mathrm{sc}}(\rp,z_0,t_s)
   +\gamma^{-2}\partial_sw,
 \label{eq:fixed-plane-K-div}\\
 \curlp\Kp^{\mathrm z}
 &=vB_z^{\mathrm{sc}}(\rp,z_0,t_s).
 \label{eq:fixed-plane-K-curl}
\end{align}
Hence,
\begin{align}
 \Kp^{\mathrm z}&=\gradp\Phi^{\mathrm z}+\ez\times\gradp\Psi^{\mathrm z},
 \label{eq:fixed-plane-Hodge}\\
 \Deltap\Phi^{\mathrm z}
 &=E_z^{\mathrm{sc}}(\rp,z_0,t_s)
   +\gamma^{-2}\partial_sw,
 &\Phi^{\mathrm z}|_{\partial\Omega_2}&=0,
 \label{eq:fixed-plane-Phi}\\
 \Deltap\Psi^{\mathrm z}&=vB_z^{\mathrm{sc}}(\rp,z_0,t_s),
 &\partial_n\Psi^{\mathrm z}|_{\partial\Omega_2}&=0,
 \label{eq:fixed-plane-Psi}\\
 &&\int_{\Omega_2}\Psi^{\mathrm z}\dd A&=0.\nonumber
\end{align}
For the same optional global-scattered convention, define
\begin{align}
 w^{\mathrm{dir}}&=\int_{-\infty}^{z_0}\overline E_z^{\mathrm{sc}}\dd z,
 &\Kp^{\mathrm{dir},\mathrm z}&=\int_{-\infty}^{z_0}\overline\Fp^{\mathrm{sc}}\dd z,
 \label{eq:fixed-plane-direct-defs}
\end{align}
The globally subtracted integrals are then
\begin{equation}
 QW_\parallel^{\mathrm{sc}}=-w^{\mathrm{dir}}-w,
 \qquad Q\Wp^{\mathrm{sc}}=\Kp^{\mathrm{dir},\mathrm z}+\Kp^{\mathrm z}.
 \label{eq:fixed-plane-complete}
\end{equation}
Thus Eqs.~\eqref{eq:fixed-plane-long-PDE}, \eqref{eq:fixed-plane-Phi}, and
\eqref{eq:fixed-plane-Psi} require a time history at one plane instead of a spatial snapshot.

The two one-sided representations of the globally subtracted wake are exactly equivalent.  From
their definitions we find
\begin{align}
 u(\rp,s)
 &=\int_{z_0-s}^{z_0}\overline E_z^{\mathrm{sc}}\dd z+w(\rp,s),
 \label{eq:u-w-relation}\\
 \Kp^{\mathrm t}(\rp,s)
 &=\int_{z_0-s}^{z_0}\overline\Fp^{\mathrm{sc}}\dd z+\Kp^{\mathrm z}(\rp,s).
 \label{eq:Kt-Kz-relation}
\end{align}
The direct parts satisfy the complementary identities
\begin{align}
 w^{\mathrm{dir}}&=u^{\mathrm{dir}}
   +\int_{z_0-s}^{z_0}\overline E_z^{\mathrm{sc}}\dd z,
 \label{eq:udir-wdir-relation}\\
 \Kp^{\mathrm{dir},\mathrm z}
 &=\Kp^{\mathrm{dir},\mathrm t}
   +\int_{z_0-s}^{z_0}\overline\Fp^{\mathrm{sc}}\dd z.
 \label{eq:Kdir-relation}
\end{align}
Hence, for $W^{\mathrm{sc}}$, a direct integral ending at $z_0-s$ and supplemented by
$(u,\Kp^{\mathrm t})$ is equal to a direct integral ending at $z_0$ and supplemented by
$(w,\Kp^{\mathrm z})$.  The two-port finite-reference wake uses the same local tail equations but a
complete finite direct integral, as defined in Sec.~\ref{sec:two-port}.

\subsection{Panofsky--Wenzel consistency}
\label{app:PW-one-sided}

We finally verify that the two one-sided decompositions satisfy the Panofsky--Wenzel relation.  This
calculation also shows why each reconstructed tail must be combined with the direct integral ending
at the same point.

For the fixed-time decomposition in Eq.~\eqref{eq:moving-complete}, the scattered-field integrals
meet at the moving point $a(s)=z_0-s$.  In the tail, differentiation of the lower limit gives
$+\overline\Fp^{\mathrm{sc}}(a,s)$, while integration of the first term in
Eq.~\eqref{eq:local-PW} gives $-\overline\E_\perp^{\mathrm{sc}}(a,s)$.  Their difference is
$v\ez\times\B_\perp^{\mathrm{sc}}(\rp,a,t_0)$ because $t[a(s),s]=t_0$.  Hence,
\begin{equation}
 \partial_s\Kp^{\mathrm t}
 =v\ez\times\B_\perp^{\mathrm{sc}}(\rp,a,t_0)-\gradp u.
 \label{eq:PW-moving-tail}
\end{equation}
The moving upper limit of the direct part gives the opposite endpoint term,
\begin{equation}
 \partial_s\Kp^{\mathrm{dir},\mathrm t}
 =-v\ez\times\B_\perp^{\mathrm{sc}}(\rp,a,t_0)-\gradp u^{\mathrm{dir}}.
 \label{eq:PW-moving-direct}
\end{equation}
Adding these equations and using Eq.~\eqref{eq:moving-complete}, we obtain
\[
 \partial_s(Q\Wp^{\mathrm{sc}})
 =-\gradp(u^{\mathrm{dir}}+u)
 =Q\gradp W_\parallel^{\mathrm{sc}}.
\]
Thus the magnetic endpoint term is caused only by the moving partition and cancels between the two
parts.

For the fixed-plane decomposition in Eq.~\eqref{eq:fixed-plane-complete}, both limits are constant.
The endpoint terms now result only from integration of
$\dd\overline\E_\perp^{\mathrm{sc}}/\dd z$ in Eq.~\eqref{eq:local-PW}.  The right tail gives
\begin{equation}
 \partial_s\Kp^{\mathrm z}
 =-\E_\perp^{\mathrm{sc}}(\rp,z_0,t_s)-\gradp w,
 \label{eq:PW-fixed-tail}
\end{equation}
while the direct integral gives
\begin{equation}
 \partial_s\Kp^{\mathrm{dir},\mathrm z}
 =\E_\perp^{\mathrm{sc}}(\rp,z_0,t_s)-\gradp w^{\mathrm{dir}}.
 \label{eq:PW-fixed-direct}
\end{equation}
Their sum is
\[
 \partial_s(Q\Wp^{\mathrm{sc}})
 =-\gradp(w^{\mathrm{dir}}+w)
 =Q\gradp W_\parallel^{\mathrm{sc}},
\]
and the electric endpoint terms cancel.

Consequently, the fixed-time tail must be paired with a direct integral ending at $a(s)$, and the
fixed-plane tail with a direct integral ending at $z_0$.  Mixing the pairs leaves the interval
between $a(s)$ and $z_0$ uncancelled.  If the moving-endpoint tail is used in the fixed-port two-port
construction, the stationary reference strip in Eq.~\eqref{eq:moving-reference-strip} converts
$a(s)$ to the prescribed port $z_2$, and Eq.~\eqref{eq:moving-two-port-trans} becomes algebraically
identical to the fixed-plane two-port result.


\begin{thebibliography}{0}%
\makeatletter
\providecommand \@ifxundefined [1]{%
 \@ifx{#1\undefined}
}%
\providecommand \@ifnum [1]{%
 \ifnum #1\expandafter \@firstoftwo
 \else \expandafter \@secondoftwo
 \fi
}%
\providecommand \@ifx [1]{%
 \ifx #1\expandafter \@firstoftwo
 \else \expandafter \@secondoftwo
 \fi
}%
\providecommand \natexlab [1]{#1}%
\providecommand \enquote  [1]{``#1''}%
\providecommand \bibnamefont  [1]{#1}%
\providecommand \bibfnamefont [1]{#1}%
\providecommand \citenamefont [1]{#1}%
\providecommand \href@noop [0]{\@secondoftwo}%
\providecommand \href [0]{\begingroup \@sanitize@url \@href}%
\providecommand \@href[1]{\@@startlink{#1}\@@href}%
\providecommand \@@href[1]{\endgroup#1\@@endlink}%
\providecommand \@sanitize@url [0]{\catcode `\\12\catcode `\$12\catcode
  `\&12\catcode `\#12\catcode `\^12\catcode `\_12\catcode `\%12\relax}%
\providecommand \@@startlink[1]{}%
\providecommand \@@endlink[0]{}%
\providecommand \url  [0]{\begingroup\@sanitize@url \@url }%
\providecommand \@url [1]{\endgroup\@href {#1}{\urlprefix }}%
\providecommand \urlprefix  [0]{URL }%
\providecommand \Eprint [0]{\href }%
\providecommand \doibase [0]{https://doi.org/}%
\providecommand \selectlanguage [0]{\@gobble}%
\providecommand \bibinfo  [0]{\@secondoftwo}%
\providecommand \bibfield  [0]{\@secondoftwo}%
\providecommand \translation [1]{[#1]}%
\providecommand \BibitemOpen [0]{}%
\providecommand \bibitemStop [0]{}%
\providecommand \bibitemNoStop [0]{.\EOS\space}%
\providecommand \EOS [0]{\spacefactor3000\relax}%
\providecommand \BibitemShut  [1]{\csname bibitem#1\endcsname}%
\let\auto@bib@innerbib\@empty
%</preamble>
\end{thebibliography}%


\begin{thebibliography}{99}

\bibitem{Weiland1983}
T.~Weiland,
``Comment on wake field computation in time domain,''
Nucl. Instrum. Methods \textbf{216}, 31--34 (1983),
\href{https://doi.org/10.1016/0167-5087(83)90327-7}{doi:10.1016/0167-5087(83)90327-7}.

\bibitem{NapolyChinZotter1993}
O.~Napoly, Y.~H.~Chin, and B.~Zotter,
``A generalized method for calculating wake potentials,''
Nucl. Instrum. Methods Phys. Res., Sect. A \textbf{334}, 255--265 (1993),
\href{https://doi.org/10.1016/0168-9002(93)90782-D}{doi:10.1016/0168-9002(93)90782-D}.

\bibitem{ZagorodnovSchuhmannWeiland2003}
I.~Zagorodnov, R.~Schuhmann, and T.~Weiland,
``Long-time numerical computation of electromagnetic fields in the vicinity of a relativistic
source,''
J. Comput. Phys. \textbf{191}, 525--541 (2003),
\href{https://doi.org/10.1016/S0021-9991(03)00329-2}{doi:10.1016/S0021-9991(03)00329-2}.

\bibitem{PanofskyWenzel1956}
W.~K.~H.~Panofsky and W.~A.~Wenzel,
``Some considerations concerning the transverse deflection of charged particles in radio-frequency
fields,''
Rev. Sci. Instrum. \textbf{27}, 967--969 (1956).

\bibitem{Zagorodnov2006}
I.~Zagorodnov,
``Indirect methods for wake potential integration,''
Phys. Rev. ST Accel. Beams \textbf{9}, 102002 (2006),
\href{https://doi.org/10.1103/PhysRevSTAB.9.102002}{doi:10.1103/PhysRevSTAB.9.102002}.

\bibitem{HenkeBruns2006}
H.~Henke and W.~Bruns,
``Calculation of wake potentials in general 3D structures,''
in \textit{Proceedings of EPAC 2006, Edinburgh, Scotland}, WEPCH110,
pp.~2170--2172 (2006).

\bibitem{ShobudaChinTakata2008}
Y.~Shobuda, Y.~H.~Chin, and K.~Takata,
``Generalized Napoly integral to compute wake potentials in axisymmetric structure,''
Phys. Rev. ST Accel. Beams \textbf{11}, 011003 (2008),
\href{https://doi.org/10.1103/PhysRevSTAB.11.011003}{doi:10.1103/PhysRevSTAB.11.011003}.

\bibitem{VaganianHenke1995}
S.~Vaganian and H.~Henke,
``The Panofsky--Wenzel theorem and general relations for the wake potential,''
Part. Accel. \textbf{48}, 239 (1995).

\bibitem{ZagorodnovWeiland2005}
I.~Zagorodnov and T.~Weiland,
``TE/TM field solver for particle beam simulations without numerical Cherenkov radiation,''
Phys. Rev. ST Accel. Beams \textbf{8}, 042001 (2005),
\href{https://doi.org/10.1103/PhysRevSTAB.8.042001}{doi:10.1103/PhysRevSTAB.8.042001}.

\bibitem{Yee1966}
K.~S.~Yee,
``Numerical solution of initial boundary value problems involving Maxwell's equations in isotropic
media,''
IEEE Trans. Antennas Propag. \textbf{14}, 302--307 (1966),
\href{https://doi.org/10.1109/TAP.1966.1138693}{doi:10.1109/TAP.1966.1138693}.

\end{thebibliography}
\end{document}